\documentclass[prl,aps,amssymb,showpacs,twocolumn]{revtex4}
\usepackage{amsmath}
\usepackage{amssymb}
\usepackage{amsthm}
\usepackage{amsfonts}
\usepackage{enumerate}
\usepackage{latexsym}
\usepackage[dvips]{graphicx}

\newcommand{\beq}{\begin{equation}}
\newcommand{\eneq}{\end{equation}}

\input{epsf}

\begin{document}

\tolerance 10000


\title{Intrinsic Spin Hall Effect in the Two Dimensional Hole Gas}

\author {B. Andrei Bernevig and Shou-Cheng Zhang}

\affiliation{  Department of Physics, Stanford University, Stanford,
CA 94305}
\begin{abstract}
\begin{center}

\parbox{17cm}{We show that two types of spin-orbit
coupling in the 2 dimensional hole gas (2DHG), with and without
inversion symmetry breaking, contribute to the intrinsic spin Hall
effect\cite{murakami2003,sinova2003}. Furthermore, the vertex
correction due to impurity scattering vanishes in both cases, in
sharp contrast to the case of usual Rashba coupling in the
electron band. Recently, the spin Hall effect in a hole doped
$GaAs$ semiconductor has been observed experimentally by
Wunderlich \emph{et al}\cite{wunderlich2004}. From the fact that
the life time broadening is smaller than the spin splitting, and
the fact impurity vertex corrections vanish in this system, we
argue that the observed spin Hall effect should be in the
intrinsic regime.}

\end{center}
\end{abstract}

\maketitle

Recent theoretical work predicts dissipationless spin currents
induced by an electric field in semiconductors with spin-orbit
coupling\cite{murakami2003,sinova2003,murakami2}. The spin current
is related to the electric field by the response equation
\begin{equation}
j_j^i = \sigma_s \epsilon_{ijk} E_k \label{spin_response}
\end{equation}
where $j_j^i$ is the current of the $i$-th component of the spin
along the direction $j$ and $\epsilon_{ijk}$ is the totally
antisymmetric tensor in three dimensions. Because both the
electric field and the spin current are even under time reversal,
the spin current could be dissipationless or intrinsic,
independent of the scattering rates. The response equation
(\ref{spin_response}) was derived by Murakami, Nagaosa and
Zhang\cite{murakami2003} for p-doped semiconductors described by
the Luttinger model of the spin-$3/2$ valence band. In another
proposal by Sinova \emph{et. al.} \cite{sinova2003}, the spin
current is induced by a in-plane electric field in the
2-dimensional electron gas (2DEG) described by the Rashba
model\cite{sinova2003}.

The spin Hall effect predicted by these recent theoretical works is
fundamentally different from the extrinsic spin Hall
\cite{d'yakonov1971,hirsch1999} effect due to Mott type of skew
scattering by impurities. The intrinsic spin Hall effect arises from
the spin-orbit coupling of the host semiconductor band, and has a
finite value in the absence of impurities. On the other hand, the
extrinsic spin Hall effect arises purely from the spin-orbit
coupling to the impurity atoms, and it is not a bulk effect like the
ordinary Hall effect. Because the extrinsic arises only from the
impurities rather than the host atoms, its magnitude is typically
many orders of magnitude smaller. The issue of impurity
contributions to the spin Hall effect has been intensively
investigated theoretically. Remarkably, Inoue \emph{et. al.}
\cite{inoue2004} calculated the vertex corrections due to impurity
scattering in the Rashba model of the electron band, and found that
the vertex correction completely cancels the spin Hall effect. Other
analytical works have obtained similar results \cite{mishchenko2004}
On the other hand, a number of numerical calculations have shown
that the spin Hall effect is independent of the disorder in the weak
disorder limit \cite{nomura2004, nikolic2004}. Currently, this
disagreement between the analytical and the numerical results is
still not settled. In a insightful paper,
Murakami\cite{murakami2004} showed that the problem of the vertex
correction does not occur in the Luttinger model of the hole band
\cite{murakami2004}. In fact, the vertex correction is identically
zero, rendering the original prediction of Ref. \cite{murakami2003}
exact in the clean limit.

Experimental observation of the spin Hall effect has been recently
reported by Kato \emph{et. al} \cite{katoscience2004} in a electron
doped sample and by Wunderlich et al in a 2DHG\cite{wunderlich2004}.
In this paper, we analyze the 2DHG experiment. In order to firmly
establish the intrinsic spin Hall effect, one needs to establish two
things. First of all, the experimental system needs to be in the
clean limit, which is the case of the 2DHG experiment, as shown in
the experimental paper\cite{wunderlich2004}. Secondly, one needs to
show that the effect is robust in the clean limit, not cancelled by
the vertex corrections due to impurities. We shall show that the
spin Hall effect in the 2DHG arises from two contributions, one from
the Luttinger Hamiltonian describing the splitting between the light
and the heavy hole bands, and one from the structural inversion
symmetry breaking (SIA) of the 2DHG band, with a spin splitting
scaling as $k^3$ \cite{winkler2000,schliemann2004}. The later form
the the spin-orbit coupling has been studied by Schliemann and
Loss\cite{schliemann2004} in connection to the intrinsic spin Hall
effect. This is different from the Rashba Hamiltonian of the 2DEG
band, where the spin splitting scales with $k$. Remarkably, we find
that the vertex correction due to impurity scattering vanishes for
both types of spin-orbit couplings in the 2DHG band, in sharp
contrast to the case of 2DEG. While the calculation details are
complicated, the intuitive reason is simple: The two types of
current vertices in the 2DHG have $p$ wave and $d$ symmetries,
respectively. When these current vertices are averaged over the $s$
wave impurity scatters, the vertex corrections vanish. These two key
facts establish a firm foundation to interpret the recent experiment
by Wunderlich et al in terms of the intrinsic spin Hall effect,
where impurities play an unessential role.

The Hamiltonian for a 2-dimensional hole gas is a sum of both
Luttinger and spin-$\vec{S} = 3/2$ SIA terms:

\begin{equation} \label{fullhamiltonian}
H = (\gamma_1 + \frac{5}{2}\gamma_2 )\frac{k^2}{2m} -
\frac{\gamma_2}{m} (\vec{k} \cdot \vec{S})^2 + \alpha (\vec{S}
\times \vec{k}) \cdot{\hat{z}}
\end{equation}
\noindent where the confinement of the well in the $z$ direction
makes the momentum be quantized on this axis. The crucial
difference between the SIA term for 2DHG and Rashba term for the
2DEG lies in the fact that $S$ is a spin $3/2$ matrix, describing
both the light (LH) and the heavy (HH) holes. For the first heavy
and light hole bands, the confinement in a well of thickness $a$
is approximated by the relation $<k_z> =0, {\langle k_z^2 \rangle}
\approx (\pi \hbar/ a)^2$. The energy eigenstates are:

\begin{widetext}
\begin{eqnarray}
&&E^{HH}_{\pm} = \frac{\gamma_1}{2m} k^2 \pm \frac{1}{2} \alpha k -
\sqrt{\alpha^2 k^2 \pm \frac{\alpha \gamma_2}{m} k (k^2 + \langle
k_z^2 \rangle) + \frac{\gamma_2^2}{m^2} (k^4 + \langle k_z^2
\rangle^2 - k^2 \langle k_z^2 \rangle)}  \nonumber \\
&& E^{LH}_{\pm} = \frac{\gamma_1}{2m} k^2 \pm \frac{1}{2} \alpha k +
\sqrt{\alpha^2 k^2 \pm \frac{\alpha \gamma_2}{m} k (k^2 + \langle
k_z^2 \rangle) + \frac{\gamma_2^2}{m^2} (k^4 + \langle k_z^2
\rangle^2 - k^2 \langle k_z^2 \rangle)}
\end{eqnarray}
\end{widetext}
\noindent  The heavy and light hole bands are split at the
$\Gamma$ point by $\Delta = 2 \gamma_2 {\langle k_z^2 \rangle}/m$
\cite{arovas1998,pala2003}. Depending on the confinement scale $a$
the Luttinger term is dominant for $a$ not too small, while the
SIA term becomes dominant for infinitely thin wells, which
correspond to high junction fields.

By expanding the above formulas for small $k << \langle k_z \rangle$
it is seen that the spin splitting of the HH bands is $~ k^3$
whereas the spin splitting of the LH bands is $~k$, in agreement
with \cite{winkler2000,winkler2002}
\begin{eqnarray}
&& E^{HH}_+  - E^{HH}_- = \frac{3}{8} \frac{ \alpha (\alpha^2 - 4
\frac{\gamma_2^2}{m^2} \langle k_z^2
\rangle)}{\frac{\gamma_2^2}{m^2} \langle k_z^2 \rangle^2} k^3 +
\mathcal{O}(k^5) \nonumber \\ && E^{LH}_{+} - E^{LH}_- = 2 \alpha k
+ \mathcal{O}(k^3)
\end{eqnarray}
\noindent Figure [1] gives a typical band structure for GaAs
($\gamma_1 = 6.92, \; \gamma_2 = 2.1$) with a $\Gamma$ point gap of
$40 meV$ and a Fermi momentum splitting of the hole band at Fermi
momentum ($0.2 nm^{-1}$) of $5meV$, which require a SIA splitting
$\alpha \approx 10^5 m/s$ .

\begin{figure}[h]
\includegraphics[scale=0.45]{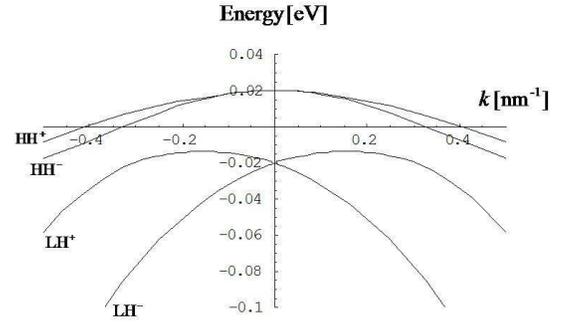}
\caption{Approximate band structure of the 2DHG ($\Delta =40 meV$,
and the spin splitting of the heavy hole band at the $k_F$ is
around $5meV$). The confinement produces a $\Gamma$ point gap
between the light and heavy hole bands, whereas the SIA produces
splitting in the previously degenerate Kramers doublets - the
heavy and light hole bands.} \label{chargedensity}
\end{figure}

 We can
expand the second term in the anisotropic Luttinger hamiltonian in
terms of Clifford algebra of Dirac $\Gamma$ matrices $\{\Gamma^a,
\Gamma^b \} = 2 \delta_{ab} I_{4 \times 4}$ ($a,b
=1,...,5$)\cite{murakami2}. Since $\sqrt{\langle k_z^2 \rangle}=0$
and $\langle k_z^2 \rangle \ne 0$ we see that the effect of
confinement renders the $d^a$'s of \cite{murakami2}:

\begin{equation}
\begin{aligned}
 (\vec{k} \cdot \vec{S})^2 &= d_a\Gamma^a; \;\; d_1  = 0, \;
d_2 =  0, \;
d_3 = - \sqrt{3}  k_x k_y,\\
d_4 &= - \frac{\sqrt{3}}{2}  (k_x^2 - k_y^2), \; d_5 =- \frac{1}{2}
(2\langle k_z^2 \rangle - k_x^2 - k_y^2)
\end{aligned}
\end{equation}
\noindent

Since calculation with the full Hamiltonian
(\ref{fullhamiltonian}) is analytically impossible, we concentrate
on different situations which maintain analytic predictability. We
now consider the case of of small junction field and neglect the
SIA term:
\begin{eqnarray}
H= \frac{\gamma_1}{2m}(k_x^2+ k_y^2 + \langle k_z^2 \rangle)+
\frac{\gamma_2}{m}d_a\Gamma^a
\end{eqnarray}
\noindent The energies are  $ E_{LH, HH} = \frac{\gamma_1}{2m}(k^2 +
\langle k_z^2 \rangle) \pm d $,
 $(d=\sqrt{d_a d_a}  = \sqrt{k^4 + \langle k_z^2 \rangle^2 - \langle
k_z^2 \rangle k^2})$.  At $k=0$ the heavy and light hole bands are
split by a gap of $\Delta E = 2 \frac{\gamma_2}{m} \langle k_z^2
\rangle \approx  2 \frac{\gamma_2}{m} (\pi \hbar/ a)^2$. In the
experiment recently reported \cite{wunderlich2004}, this energy
gap is of order $\Delta  E = 40 meV$, which corresponds to an $a=
8.3 nm $ thick quantum well. The value quoted in the experiment is
roughly $3-4 nm$, making our simplistic prediction rather
accurate.

We would like to compute the response of spin current $J_{i}^l =
\frac{1}{2} \{ S^l, \frac{\partial H }{\partial k_j} \}$ to an
electric current $J_j= \frac{
\partial H }{
\partial k_j}$:
\begin{equation}\label{responsefunction}
Q^{l}_{ij} (i \nu_m) = -\frac{1}{V} \int_0^{\beta} \langle T
J^{l}_i(u) J_j \rangle  e^{i \nu_m u} du
\end{equation}
\noindent The spin conductance is then defined as $\sigma^{l}_{ij}
= \lim_{\omega \rightarrow 0} \frac{Q^l_{ij}(\omega)}{ - i\omega}$
and gives:
\begin{equation}
 \sigma^l_{ij}  =\frac{1}{ V } \sum_{k} \frac{n^F_+ - n^F_-}{ d^3}
 \eta^l_{ab} \left[2 \frac{m}{\gamma_2} d_b \frac{\partial d_a}{\partial k_j}
\frac{\partial \varepsilon}{\partial k_i} + \epsilon_{abcde} d_e
\frac{\partial d_c}{\partial k_i} \frac{\partial d_d}{\partial k_j}
\right]
\end{equation}
\noindent  where $\eta^l_{ab}$ is a tensor defined in
\cite{murakami2}, $n^F_\pm$ are the Fermi functions of the two bands
and $\varepsilon = \frac{\gamma_1}{2m}(k_x^2+ k_y^2 + \langle k_z^2
\rangle)$ is the kinetic energy. The last term is the conserved spin
conductance \cite{murakami2} (which represents the response of the
spin projected onto the HH and LH bands \cite{murakami2}), whereas
the first term is the contribution of the non-conserved part of the
spin. Upon momentum integration, since motion along the $z$ axis is
prohibited due to confinement, the only non-zero components are
$\sigma^{3}_{12}=-\sigma^{3}_{21}$ which yield:
\begin{widetext}
\begin{equation}
\sigma^3_{12} =\frac{1}{4 \pi} \left( \frac{3}{2}
\frac{\gamma_1}{2\gamma_2}\left[ \frac{2 (k^2 + \langle k_z^2
\rangle)}{3 \sqrt{k^4 + \langle k_z^2 \rangle^2 - \langle k_z^2
\rangle k^2}} - \ln[2 \sqrt{k^4 + \langle k_z^2 \rangle^2 - \langle
k_z^2 \rangle k^2} + 2 k^2 - \langle k_z^2 \rangle] \right]     +
\frac{ 2 \langle k_z^2 \rangle - k^2}{\sqrt{k^4 + \langle k_z^2
\rangle^2 - \langle k_z^2 \rangle k^2}} \right)^{k=k_{LH}}_{k =
k_{HH }}
\end{equation}
\end{widetext}
\noindent   where $k_{LH}, k_{HH}$ are the fermi momenta of the
light and heavy hole bands. For the experimental data
\cite{wunderlich2004}, the light hole band is fully occupied, so
$k_{LH} = 0$ while $\sqrt{\langle k_z^2 \rangle} = 3.7 \times
10^{-26} kg \; m/s$ and $k_{HH} = 3 \times 10^{-26} kg \; m/s$. The
first two terms are due to the non-conserved spin and for $GaAs$,
$\sigma^{3 (noncons)}_{1,2} = \frac{0.7}{8 \pi}$. The last term is
the conserved spin conductance $\sigma^{3(cons)}_{12} = 0.6 \times
\frac{1}{4 \pi}$. The total spin conductance is therefore
$\sigma^3_{12} = \frac{1.9}{8 \pi}$, in good agreement with the
numerical estimate in \cite{wunderlich2004}. In the case of infinite
confinement, $\sqrt{\langle k_z^2 \rangle} \rightarrow \infty$ the
spin conductance from the Luttinger term vanishes, as it should,
since we would then enter a SIA dominated regime.

We now investigate the effect of disorder on the Luttinger spin Hall
conductance. In particular, we want to find out the vertex
correction. The free Green's function in our system is defined as
$G_0({\bf{k}}, E)= [E-H]^{-1}$:
\begin{equation}
G_0({\bf{k}}, i \omega_n) =\frac{ i \omega_n  - \epsilon({\bf k}) +
\frac{\gamma_2}{m}d_a \Gamma_a}{(i \omega_n  - \epsilon({\bf k}))^2
- \gamma_2^2 d^2/m^2}
\end{equation}
\noindent We model the disorder as randomly distributed,
spin-independent identical defects $V({\bf{r}} )= u \sum_i
\delta({\bf{r}} -{\bf{R}}_i)$. In the Born approximation, the
self-energy is related to the free Green function $\Sigma(i
\omega_n) = n_{imp} u^2 \int \frac{d {\bf{k}}}{(2 \pi)^2}
G_0({\bf{k}}, i \omega_n) $ Since $\int d{\bf k} d_3 = \int d{\bf k}
d_4 =0$, the self energy is an isotropic function of $\vec{k}$:
\begin{equation}
\Sigma(i \omega_n)  = n_{imp} u^2 \int \frac{d {\bf{k}}}{(2 \pi)^2}
\frac{ i \omega_n  - \epsilon({\bf k}) + \frac{\gamma_2}{m}d_5
\Gamma_5}{(i \omega_n  - \epsilon({\bf k}))^2 - \gamma_2^2 d^2/m^2}
\end{equation}
\noindent where $d_5=- \frac{1}{2} (2\langle k_z^2 \rangle - k^2)$.
This is different from the bulk Luttinger case, where the $d_5(k)$
integral over $\vec{k}$ vanishes as well, but the difference is not
essential. The full impurity Green function is $G({\bf {k}},
i\omega_n) = G_0({\bf{k}}, i \omega_n + \Sigma(i\omega_n))$. The
current vertex satisfies a Bethe-Salpeter equation similar to
[\ref{bethesalpeter}]. Similar to the case of \cite{murakami2004},
since the Green's function is an even function of the in-plane total
momentum, while the charge current operator $V_j =\partial H /
\partial k_j$ is momentum-odd in the components $k_j$ (because the Hamiltonian $H$ is even in $\vec{k}$,
it turns out that the free vertex cancels
\begin{equation}
\int \frac{d {\bf k}}{(2 \pi)^2} G({\bf{k}}, i\omega_n)
V_j({\bf{k}}) G({\bf{k}}, i\omega_n- i \nu_m) = 0
\end{equation}
\noindent And hence the vertex correction which is an iterative
function of the free vertex vanishes as well \cite{murakami2004}.

To see the effect of a very small SIA splitting ($\alpha k_F <<
\Delta, \;\; \alpha k_F << \hbar / \tau$) on the Luttinger spin
Hall conductance, we treat it in perturbation theory. The
calculation is long and not particularly revealing, so we just
give the result. Due to the fact that the Luttinger current
operator is odd in $\vec{k}$ while the SIA current operator is a
constant matrix, the first order contribution in $\alpha$
vanishes. The first nonzero contribution is of order $\alpha^2$.

We now turn to the opposite case, of strongly confined quantum
wells, in which the SIA term is likely to dominate. We model the
system by a Gamma point gap $\Delta $ plus a spin $3/2$ Rashba term
$\alpha (\vec{k} \times \vec{S} ) \hat{z}$ \cite{bernevigprb2}. We
compute the spin conductance and expand it in terms of the ratio
between the SIA spin splitting and the $\Gamma$ point gap,
$\frac{\alpha k_f}{ \Delta} < 1$. The spin conductance gets a
contribution from the HH band:
\begin{equation} \label{heavyhole1}
\sigma^{3 (HH)}_{12}  = \frac{9}{8 \pi} (1+ \frac{\alpha^2 m_{HH}}{2
\Delta})
\end{equation}
\noindent For infinitely thin quantum wells, $\Delta \rightarrow
\infty$, the HH spin conductance is $9 / 8 \pi$ which is (besides a
re-scaling factor of $2$ in the spin current definition), the same
as that obtained in \cite{schliemann2004} who studied the hole gas
by starting with the effective HH Hamiltonian directly. The second
term in Eq[\ref{heavyhole1}] is the first order finite thickness
correction. If the Fermi level is low enough, there is also a
light-hole band contribution to the spin conductance of order $
\sigma^{3 (LH)}_{12} = \frac{1}{8 \pi} (1+ \frac{3\alpha^2 m_{LH}}{2
\Delta})$. The vertex correction for a spin-$3/2$ Rashba-like system
has been computed in $3D$ perturbatively to first order in $\alpha$
and was found to be finite \cite{bernevigvafek2004}. We now compute
it exactly for the $2D$ case in the heavy hole band.

Since working with spin $3/2$ matrices is cumbersome and we do not
need the LH states as they are fully filled \cite{wunderlich2004},
we now project the system onto the heavy hole states and work with
the truncated Hamiltonian \cite{winkler2000,schliemann2004}:
\begin{equation}
H =\frac{k^2}{2 m} + \beta (k_-^3 \sigma_+ - k_+^3 \sigma_-)
\end{equation}
\noindent which becomes exact in the limit of very confined quantum
wells. The spin Hall conductance in the disorder-free case is $9 /8
\pi$, as previously obtained in \cite{schliemann2004} and as
obtained above as the limit of strongly confined quantum wells. The
Hamiltonian can also be expressed:
\begin{equation}
H = \frac{k^2}{2 m} + \lambda_i(k) \sigma_i, \;\;\; i=x,y
\end{equation}
\noindent Where $\lambda_1 = \beta k_y(3 k_x^2 - k_y^2)$ and
$\lambda_2= \beta k_x(3 k_y^2 - k_x^2)$, with $\beta$ a constant.
Let $\lambda(k)=\sqrt{\lambda_i \lambda_i}$. The Fermi sphere is
isotropic since the energy levels are $ E_\pm = \frac{k^2}{2 m} \pm
\lambda$. The disorder-free Green function is:
\begin{equation}
G_0 ({\bf k} , i \omega_n) = \frac{1}{2} \sum_{s = \pm} \frac{1 + s
\hat{\lambda}_i \sigma_i}{i \omega_n - E_s}
\end{equation}
\noindent where $\hat{\lambda}_i = \lambda_i / \lambda$. The self
energy for s-wave scattering of electrons becomes a
state-independent constant (not a matrix) $\Sigma (i \omega_n) =
n_{imp} u^2 \int \frac{d {\bf k}}{(2 \pi)^2} G_0({\bf k}, i
\omega_n) $ where $n_{imp}$ is the density of impurities while $u$
is the impurity potential strength . Since $\lambda_i$ are odd
functions of the momentum components $k_i$, the integral $\int d
{\bf k} f(k) \lambda_i = 0$ where $f(k)$ is any isotropic function
of $k$. Since the spin orbit coupling small (much smaller than the
Fermi energy), the density of states at zero order is a constant $D
=m/2 \pi \hbar^2$ while the $\alpha k^3$ term in the Hamiltonian
contributes to with only a first order correction. The full Green
function in the presence of impurities is $G({\bf k}, i \omega_n) =
G_0 ({\bf k}, i\omega_n + \Sigma (i \omega_n))$. The spin dependent
part of the charge current operator $V_j(k) =
\partial H/ \partial k_j$ turns out to have $d$-wave symmetry (for
example, the spin dependent part of the $V_x$ operator reads $6
\beta k_x k_y \sigma_x + 3 \beta(k_y^2 -k_x^2) \sigma_y$ ) and it
vanishes when integrated over the isotropic Fermi surface. This is
the deep intuitive reason as to why the vertex correction cancels in
this case, as we rigorously show below. By contrast, in the
electron-band Rashba case, the spin-dependent part of the charge
operator is a constant. The current vertex function $K_j ({\bf{k}} ,
i \omega_n, i \nu_m) =\langle G ({\bf k}, i \omega_n) V_j(k) G({\bf
k}, i \omega_n + i \nu_m) \rangle$ is a matrix function that does
not commute with either the charge current operator or the Green's
function. $\langle ... \rangle $  is an impurity average. It
satisfies the Bethe-Salpeter equation:
\begin{widetext}
\begin{equation} \label{bethesalpeter}
K_j ({\bf{k}} , i \omega_n, i \nu_m) =  G({\bf{k}}, i\omega_n)
V_j({\bf{k}}) G({\bf{k}}, i\omega_n- i \nu_m) + n_{imp}u^2
G({\bf{k}}, i\omega_n ) \int \frac{d {\bf{q}}}{(2 \pi)^2} K_j
({\bf{q}} , i \omega_n, i \nu_m) G({\bf{k}}, i\omega_n- i \nu_m)
\end{equation}
\end{widetext}
\noindent Integrating both the right and the left hand side over the
momentum ${\bf{k}}$, we see that the vertex correction $\Delta
V_j(i\omega_n, i \nu_m) = \int \frac{d {\bf{q}}}{(2 \pi)^2} K_j
({\bf{q}} , i \omega_n, i \nu_m) $ satisfies:
\begin{eqnarray}
&& \Delta V_j = \int \frac{d {\bf k}}{(2 \pi)^2} G({\bf{k}},
i\omega_n) V_j({\bf{k}}) G({\bf{k}}, i\omega_n- i \nu_m) + \nonumber
\\ && +n_{imp}u^2 \int \frac{d {\bf k}}{(2 \pi)^2} G({\bf{k}},
i\omega_n ) \Delta V_j G({\bf{k}}, i\omega_n- i \nu_m)
\end{eqnarray}
\noindent Since the vertex correction $\Delta V_j (i \omega_n, i
\nu_m)$ is a $2\times 2 $ matrix, it can be decomposed in the basis
of the identity matrix and the $3$ pauli matrices:
\begin{equation}
\Delta V_j (i \omega_n, i \nu_m) = \sum_{\mu = 0}^3 \Lambda^\mu_j (i
\omega_n, i \nu_m) \sigma^\mu, \;\;\; \mu = 0,1,2,3
\end{equation}
\noindent where $\sigma^0 = I_{2 \times 2}$, the identity matrix,
and $\sigma^{1,2,3}$ are the $3$ Pauli matrices. The $\Lambda^\mu_j
(i \omega_n, i \nu_m)$ are scalars. By introducing the decomposition
in the vertex equation, multiplying to the left of both sides of the
equal by a $\sigma_\alpha$ matrix and taking the trace of the above
equation, we obtain:
\begin{eqnarray}\label{vertexexpanded}
&& 2 \Lambda^\nu_j  = A^\nu_j(i \omega_n, i \nu_m) + \sum_{\mu =
0}^3 \Lambda^\mu_j    M^{\nu \mu}(i \omega_n, i \nu_m) \nonumber \\
&& M^{\nu \mu}  = n_{imp} u^2 \int \frac{d {\bf k}}{(2 \pi)^2} Tr[
\sigma^\nu G({\bf{k}}, i\omega_n ) \sigma^\mu G({\bf{k}}, i\omega_n-
i \nu_m)]  \nonumber \\
&& A^\nu_j  = \int \frac{d {\bf k}}{(2 \pi)^2} Tr[\sigma^\nu
G({\bf{k}}, i\omega_n) V_j({\bf{k}}) G({\bf{k}}, i\omega_n- i
\nu_m)]
\end{eqnarray}
\noindent By expanding and evaluating $M^{\nu \mu}$ (this uses the
observation that $\int d{\bf k} \lambda^i(k) \lambda^j (k) \sim
\delta_{ij}$ as well as  $G^R_s G^A_s = \frac{2 \pi \tau}{\hbar}
\delta(\epsilon - E_s)$, where $R, A$ stand for the retarded and
advanced Green's functions, and $\tau = \hbar^3/ n_{imp} u^2 m$) we
observe that it is diagonal in $\mu, \nu$, that is $M^{\nu \mu} =
\delta_{\nu \mu}$. Expanding the traces in
Eq.(\ref{vertexexpanded}), and since $\lambda_0(k) = \lambda_3(k) =
0$ it is easy to observe that (after azimuthal integration) $A^0_j
(i \omega_n, i \nu_m)= A^3_j (i \omega_n, i \nu_m)=0$ and hence the
vertex corrections $\Lambda^0_j(i \omega_n, i \nu_m) = \Lambda^3_j
(i \omega_n, i \nu_m) =0$. We now have for the vertex correction $
\Lambda^\nu_j (i \omega_n, i \nu_m) = A^\nu_j(i \omega_n, i \nu_m),
\;\;\; \nu=1,2$ and $ j =x,y $ where after expanding the traces:
\begin{equation} \label{theas}
A^\nu_j =  \sum_{s, s' = \pm } \int \frac{d {\bf k}}{(2
\pi)^2}\frac{(s+s')\frac{k_j}{m} \hat{\lambda}_\nu + 2 s s'
\hat{\lambda}_\nu \frac{\partial \lambda}{ \partial k_j} +
(1-ss')\frac{\partial \lambda_\nu }{
\partial k_j}}{2(z-E_s)(z'- E_{s'})}
\end{equation}
\noindent with $z = i\omega_n + \Sigma(i\omega_n)$, $z'=  i\omega_n
- i\nu_m + \Sigma(i\omega_n- i \nu_m)$. We now compute this for $\nu
=1$, the case $\nu=2$ being identical. Let $j=1$ and we find for the
numerator of the integrand in Eq(\ref{theas}):
\begin{equation}
  \frac{k_y k_x (3 k_x^2
- k_y^2)}{k^3} \left[ (s+s') \frac{1}{m} + 6 ss' \beta k\right] + 6
\beta k_x k_y (1- ss')
\end{equation}
\noindent Upon integration over $d {\bf k}$ the above expression
vanishes due to the integral over the azimuthal angle and hence
$A^1_1 (i \omega_n, i \nu_m) = 0$.  For the case $\nu = 1, j=2$, the
numerator of Eq(\ref{theas}) gives:
\begin{equation}
 \frac{k_y^2 (3 k_x^2 -
k_y^2)}{k^3} \left[ (s+s') \frac{1}{m} + 6 ss' \beta k\right]
 + 3 \beta (k_x^2 - k_y^2) (1- ss')
\end{equation}
\noindent which also vanishes upon azimuthal angle integration
$A^1_2 (i \omega_n, i \nu_m) = 0$. In an identical way all the
components of the vertex correction tensor vanish.

We have analyzed the spin Hall transport in the case of a
two-dimensional hole gas. We showed that for relative weak
confinement the spin-Hall conductance is of Luttinger type and is
equal to roughly $1.9 e/8\pi$ for the value of parameters in
\cite{wunderlich2004}. For strongly confined quantum wells, the
system is dominated by a structural inversion asymmetry term of
spin-$3/2$ SIA-type. The spin conductance for this system is $9
e/8 \pi$ plus a correction dependent on the quantum-well size. We
perform the full vertex correction and show that it vanishes for
both Luttinger and SIA cases. This is in striking contrast to the
$k$-linear Rashba case, where the vertex correction is of the same
magnitude and of opposite sign to the spin orbit coupling
strength. Coupled with the fact that the life time broadening is
smaller than the spin splitting, we hence conclude that the spin
Hall effect observed in \cite{wunderlich2004} should be in the
intrinsic regime. The dissipationless spin Hall conductance can be
systematically determined by extrapolating the ratio of life time
broadening and the spin splitting to zero.

 We would like thank J. Sinova and J. Wunderlich for useful
discussions. B.A.B. acknowledges support from the Stanford Graduate
Fellowship Program. This work is supported by the NSF under grant
numbers DMR-0342832 and the US Department of Energy, Office of Basic
Energy Sciences under contract DE-AC03-76SF00515.


\begin{thebibliography}{19}
\expandafter\ifx\csname
natexlab\endcsname\relax\def\natexlab#1{#1}\fi
\expandafter\ifx\csname bibnamefont\endcsname\relax
  \def\bibnamefont#1{#1}\fi
\expandafter\ifx\csname bibfnamefont\endcsname\relax
  \def\bibfnamefont#1{#1}\fi
\expandafter\ifx\csname citenamefont\endcsname\relax
  \def\citenamefont#1{#1}\fi
\expandafter\ifx\csname url\endcsname\relax
  \def\url#1{\texttt{#1}}\fi
\expandafter\ifx\csname urlprefix\endcsname\relax\def\urlprefix{URL
}\fi \providecommand{\bibinfo}[2]{#2}
\providecommand{\eprint}[2][]{\url{#2}}

\bibitem[{\citenamefont{Murakami et~al.}(2003)\citenamefont{Murakami, Nagaosa,
  and Zhang}}]{murakami2003}
\bibinfo{author}{\bibfnamefont{S.}~\bibnamefont{Murakami}},
  \bibinfo{author}{\bibfnamefont{N.}~\bibnamefont{Nagaosa}}, \bibnamefont{and}
  \bibinfo{author}{\bibfnamefont{S.}~\bibnamefont{Zhang}},
  \bibinfo{journal}{Science} \textbf{\bibinfo{volume}{301}},
  \bibinfo{pages}{1348} (\bibinfo{year}{2003}).

\bibitem[{\citenamefont{\text{J. Sinova} \emph{et. al.}}(2004)}]{sinova2003}
\bibinfo{author}{\bibnamefont{\text{J. Sinova} \emph{et. al.}}},
  \bibinfo{journal}{Phys. Rev. Lett.} \textbf{\bibinfo{volume}{92}},
  \bibinfo{pages}{126603} (\bibinfo{year}{2004}).

\bibitem[{\citenamefont{\text{J. Wunderlich }
  \emph{et.al.}}()}]{wunderlich2004}
\bibinfo{author}{\bibnamefont{\text{J. Wunderlich } \emph{et.al.}}},
  \bibinfo{howpublished}{cond-mat/0410295}.

\bibitem[{\citenamefont{Murakami et~al.}(2004)\citenamefont{Murakami, Nagaosa,
  and Zhang}}]{murakami2}
\bibinfo{author}{\bibfnamefont{S.}~\bibnamefont{Murakami}},
  \bibinfo{author}{\bibfnamefont{N.}~\bibnamefont{Nagaosa}}, \bibnamefont{and}
  \bibinfo{author}{\bibfnamefont{S.}~\bibnamefont{Zhang}},
  \bibinfo{journal}{Phys. Rev. B} \textbf{\bibinfo{volume}{69}},
  \bibinfo{pages}{235206} (\bibinfo{year}{2004}).

\bibitem[{\citenamefont{M.I.D'yakonov}(1971)}]{d'yakonov1971}
\bibinfo{author}{\bibfnamefont{V.~P.} \bibnamefont{M.I.D'yakonov}},
  \bibinfo{journal}{Phys. Lett. A} \textbf{\bibinfo{volume}{35}},
  \bibinfo{pages}{459} (\bibinfo{year}{1971}).

\bibitem[{\citenamefont{Hirsch}(1999)}]{hirsch1999}
\bibinfo{author}{\bibfnamefont{J.}~\bibnamefont{Hirsch}},
  \bibinfo{journal}{Phys. Rev. Lett.} \textbf{\bibinfo{volume}{83}},
  \bibinfo{pages}{1834} (\bibinfo{year}{1999}).

\bibitem[{\citenamefont{Inoue et~al.}(2004)\citenamefont{Inoue, Bauer, and
  Molenkamp}}]{inoue2004}
\bibinfo{author}{\bibfnamefont{J.}~\bibnamefont{Inoue}},
  \bibinfo{author}{\bibfnamefont{G.}~\bibnamefont{Bauer}}, \bibnamefont{and}
  \bibinfo{author}{\bibfnamefont{L.}~\bibnamefont{Molenkamp}},
  \bibinfo{journal}{Phys. Rev. B} \textbf{\bibinfo{volume}{70}},
  \bibinfo{pages}{041303} (\bibinfo{year}{2004}).

\bibitem[{\citenamefont{Mishchenko et~al.}()\citenamefont{Mishchenko, Shytov,
  and Halperin}}]{mishchenko2004}
\bibinfo{author}{\bibfnamefont{E.}~\bibnamefont{Mishchenko}},
  \bibinfo{author}{\bibfnamefont{A.}~\bibnamefont{Shytov}}, \bibnamefont{and}
  \bibinfo{author}{\bibfnamefont{B.}~\bibnamefont{Halperin}},
  \bibinfo{howpublished}{cond-mat/0406730}.

\bibitem[{\citenamefont{\text{K. Nomura} \emph{et. al.}}()}]{nomura2004}
\bibinfo{author}{\bibnamefont{\text{K. Nomura} \emph{et. al.}}},
  \bibinfo{howpublished}{cond-mat/0407279}.

\bibitem[{\citenamefont{Nikolic et~al.}()\citenamefont{Nikolic, Zarbo, and
  Sauma}}]{nikolic2004}
\bibinfo{author}{\bibfnamefont{B.}~\bibnamefont{Nikolic}},
  \bibinfo{author}{\bibfnamefont{L.}~\bibnamefont{Zarbo}}, \bibnamefont{and}
  \bibinfo{author}{\bibfnamefont{S.}~\bibnamefont{Sauma}},
  \bibinfo{howpublished}{cond-mat/0408693}.

\bibitem[{\citenamefont{Murakami}(2004)}]{murakami2004}
\bibinfo{author}{\bibfnamefont{S.}~\bibnamefont{Murakami}},
  \bibinfo{journal}{Phys. Rev. B} \textbf{\bibinfo{volume}{69}},
  \bibinfo{pages}{241202(R)} (\bibinfo{year}{2004}).

\bibitem[{\citenamefont{\text{Y. Kato} \emph{et. al.}}()}]{katoscience2004}
\bibinfo{author}{\bibnamefont{\text{Y. Kato} \emph{et. al.}}},
  \bibinfo{howpublished}{Science, 11 Nov 2004 (10.1126/science.1105514)}.

\bibitem[{\citenamefont{Winkler}(2000)}]{winkler2000}
\bibinfo{author}{\bibfnamefont{R.}~\bibnamefont{Winkler}},
  \bibinfo{journal}{Phys. Rev. B} \textbf{\bibinfo{volume}{62}},
  \bibinfo{pages}{4245} (\bibinfo{year}{2000}).

\bibitem[{\citenamefont{Schliemann and Loss}()}]{schliemann2004}
\bibinfo{author}{\bibfnamefont{J.}~\bibnamefont{Schliemann}} \bibnamefont{and}
  \bibinfo{author}{\bibfnamefont{D.}~\bibnamefont{Loss}},
  \bibinfo{howpublished}{cond-mat/0405436}.

\bibitem[{\citenamefont{Arovas and Geller}(1998)}]{arovas1998}
\bibinfo{author}{\bibfnamefont{D.}~\bibnamefont{Arovas}} \bibnamefont{and}
  \bibinfo{author}{\bibfnamefont{Y.}~\bibnamefont{Geller}},
  \bibinfo{journal}{Phys. Rev. B} \textbf{\bibinfo{volume}{57}},
  \bibinfo{pages}{12302} (\bibinfo{year}{1998}).

\bibitem[{\citenamefont{\text{M.G. Pala} \emph{et. al.}}()}]{pala2003}
\bibinfo{author}{\bibnamefont{\text{M.G. Pala} \emph{et. al.}}},
  \bibinfo{howpublished}{cond-mat/0307354}.

\bibitem[{\citenamefont{\text{R. Winkler} \emph{et. al}}(2002)}]{winkler2002}
\bibinfo{author}{\bibnamefont{\text{R. Winkler} \emph{et. al}}},
  \bibinfo{journal}{Phys. Rev. B} \textbf{\bibinfo{volume}{65}},
  \bibinfo{pages}{155303} (\bibinfo{year}{2002}).

\bibitem[{\citenamefont{Bernevig and Zhang}()}]{bernevigprb2}
\bibinfo{author}{\bibfnamefont{B.}~\bibnamefont{Bernevig}} \bibnamefont{and}
  \bibinfo{author}{\bibfnamefont{S.}~\bibnamefont{Zhang}},
  \bibinfo{howpublished}{in preparation}.

\bibitem[{\citenamefont{Bernevig and O.Vafek}()}]{bernevigvafek2004}
\bibinfo{author}{\bibfnamefont{B.}~\bibnamefont{Bernevig}} \bibnamefont{and}
  \bibinfo{author}{\bibnamefont{O.Vafek}},
  \bibinfo{howpublished}{cond-mat/0406153}.

\end{thebibliography}
\end{document}